\documentclass[%
 reprint,
 amsmath,amssymb,
 aps,
]{revtex4-2}

\usepackage{graphicx}% Include figure files
\usepackage{dcolumn}% Align table columns on decimal point
\usepackage{bm}% bold math
\usepackage{hyperref}

\begin{document}

\title{Quasiparticle dynamics in a superconducting qubit irradiated by a localized infrared source}

\author{R. Benevides$^{1,2}$}\thanks{These authors contributed equally to the work}
\author{M. Drimmer$^{1,2}$}\thanks{These authors contributed equally to the work}
\author{G. Bisson$^{1,2}$}%
\author{F. Adinolfi$^{1,2}$}%
\author{U. v. Lüpke$^{1,2}$}
\author{H. M. Doeleman$^{1,2}$}%
\author{G. Catelani$^{3,4}$}
\author{Y. Chu$^{1,2}$}%
\affiliation{$^{1}$Department of Physics, ETH Zürich, 8093 Zürich, Switzerland}
\affiliation{$^{2}$Quantum Center, ETH Zürich, 8093 Zürich, Switzerland}
\affiliation{$^{3}$Jülich Aachen Research Alliance (JARA) Institute for Quantum Information (PGI-11), Forschungszentrum Jülich, Jülich 52425, Germany}
\affiliation{$^{4}$Quantum Research Center, Technology Innovation Institute, Abu Dhabi 9639, United Arab Emirates}

\date{\today}

\begin{abstract}
A known source of decoherence in superconducting qubits is the presence of broken Cooper pairs, or quasiparticles. These can be generated by high-energy radiation, either present in the environment or purposefully introduced, as in the case of some hybrid quantum devices. Here, we systematically study the properties of a transmon qubit under illumination by focused infrared radiation with various powers, durations, and spatial locations. Despite the high energy of incident photons, our observations agree well with a model of low-energy quasiparticle dynamics dominated by trapping. This technique can be used for understanding and potentially mitigating the effects of high-energy radiation on superconducting circuits with a variety of geometries and materials.  

\end{abstract}

\maketitle

The quantum coherence of superconducting (SC) circuits has steadily improved in recent years \cite{Paik2011, Siddiqi2021, place2021,ganjam2023}. This has enabled them to not only become one of the leading quantum information processing platforms, but also a crucial ingredient in a variety of hybrid devices aiming to combine superconducting circuits with other quantum systems~\cite{clerk2020hybrid,chu2020perspective}. However, understanding and mitigating the sources of decoherence in SC circuits is still an important effort in this field, particularly when considering hybrid systems where new components and degrees of freedom are introduced.

One significant decoherence mechanism is the breaking of Cooper pairs in the superconductor, creating so-called \textit{quasiparticles}~\cite{Glazman2021,Arutyunov2018,deVisser2011}. The presence of quasiparticles can lead to both energy relaxation and dephasing of qubits \cite{catelani2011, Martinis2009, lenander2011measurement, CatelaniPRB2012,Aumentado2004,Shaw2008}. The first of these effects was studied using controlled injection of quasiparticles near Josephson junctions with microwave drives \cite{Wang2014,liu2022quasiparticle}. It has also been shown that quasiparticles generated by high-energy particles, such as high-energy photons or the products of radioactive decay~\cite{vepsalainen2020impact}, can even result in correlated errors between multiple qubits~\cite{wilen2021correlated, McEwen2022}, which is especially detrimental for quantum error correction protocols~\cite{martinis2021saving}. 

In an effort to mitigate these detrimental effects, previous studies have focused on shielding SC circuits from environmental high-energy radiation~\cite{Corcoles2011,Barends2011,vepsalainen2020impact, cardani2021reducing}. However, quasiparticle densities measured are still orders of magnitude higher than predicted at thermal equilibrium~\cite{serniak2018hot,Pan2022}.  In addition, hybrid devices such as microwave-optical quantum transducers require the introduction of a large number of optical photons near SC circuits, which can lead to significant additional decoherence that degrades the transducer's performance~\cite{Mirhosseini2020, Delaney2022,chu2020perspective,xu2023lightinduced}.

In this work, we use a complementary approach that introduces, on purpose, high-energy radiation in the vicinity of a SC circuit. Specifically, we direct a telecom-band laser beam onto a three-dimensional (3D) aluminum transmon qubit device~\cite{Paik2011} in a controllable manner in order to systematically investigate the impact of the incident photons on qubit properties. We are able to vary not only the power and duration but also the location of the focused laser beam relative to the qubit. With this method, we demonstrate unprecedented temporal and spatial control over the generation of quasiparticles. We show that the transmon's excited state lifetime is modified by the incident light, but recovers about an order of magnitude faster than in previous experiments with microwave induced quasiparticles \cite{Wang2014}. We find that the recovery dynamics of the qubit coherence follows a universal dependence on the incident energy and that it is well described by a model of quasiparticle population decay through trapping and recombination \cite{Wang2014}. Moreover, measurements of $T_2^*$ allow us to test the contribution of quasiparticle-induced pure dephasing to the qubit decoherence, which, in agreement with theoretical models~\cite{CatelaniPRB2012,catelani2014parity}, is smaller than the contribution from induced energy decay.  

\section{Experimental design}

\begin{figure}[!t]
\includegraphics{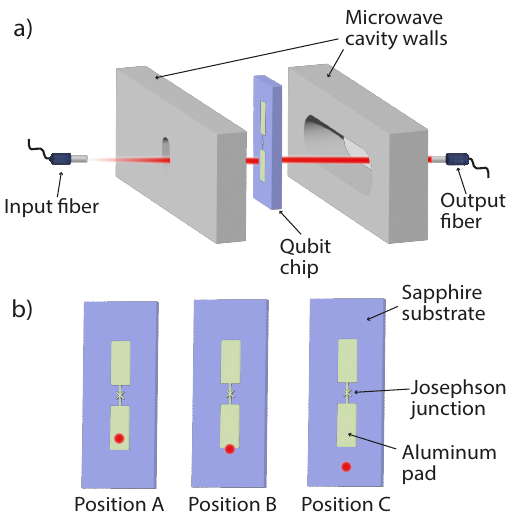}
\caption{\label{fig:setup} Schematic of the setup. (a) An infrared laser beam enters its microwave cavity via holes in the walls, reaching the qubit chip. Transmitted light is sent to a collimator (output fiber) and used to construct a low-resolution picture of the qubit. (b) Diagram of the three positions of the laser beam used in this work, labeled A, B, and C.}
\end{figure}

Our device consists of an aluminum transmon qubit on a sapphire substrate situated in a 3D aluminum microwave cavity at the base plate of a dilution refrigerator operating around 10 mK. 0.5 mm wide holes in the walls of the cavity allow laser light to enter and exit the cavity, as seen in Fig.~\ref{fig:setup}(a). Light from a 1550-nm-wavelength laser is guided into the dilution refrigerator via an optical fiber and focused into the cavity to a beam waist of 47 $\mu$m at the plane of the qubit. Variable optical attenuators (VOAs) and a fiber-coupled acousto-optic modulator (AOM) are used to control the power and duration of the laser pulses. At the base plate of the dilution refrigerator, the fiber is glued to an input lens and mounted on a motorized tip-tilt stage that allows us to change the position of the focused laser spot with respect to the qubit. A multimode fiber coupler with a high numerical aperture is used to collect the light that exits on the other side of the microwave cavity. This light is then directed out of the fridge and measured with a photodiode. By changing the position of the laser beam and measuring the transmitted light, we can take low-resolution images that we use to position the laser beam relative to the qubit (see Supplementary Material~\cite{SM}, sections~\ref{sec:SI-details} and~\ref{sec:SI-imaging}). 

We performed measurements with three different laser beam locations during a single cooldown, as shown in Fig.~\ref{fig:setup}(b): In position A, the laser is focused directly onto the metal pad of the qubit. In position B, the laser is focused on the bottom edge of the qubit pad. In position C, the laser is focused on the transparent sapphire substrate, approximately $200~\mu$m below the qubit. 

The typical individual photon energy ($h\times 193.4$~THz) in the infrared regime is orders of magnitude larger than the superconducting bandgap $2\Delta$  of aluminum ($\Delta = h\times 46.9$~GHz, for thin-film aluminum~\cite{Paik2011}). An infrared photon impinging on the device creates high-energy excitations in the superconducting film and substrate, which are converted into a large number of low-energy quasiparticles through a cascading microscopic process consisting of scattering interactions involving phonons, electrons, and other quasiparticles~\cite{Kozorev2000}. The number of low-energy quasiparticles then decreases over time due to two main processes: First, quasiparticles can be individually trapped at the edges of the superconductor or in vortices caused by residual magnetic fields. Second, these electronic excitations can recombine into Cooper pairs, releasing their energy mostly via phonon creation and returning to the superconducting condensate~\cite{kaplan1976quasiparticle}. 

The light-induced quasiparticle density in the vicinity of the transmon's Josephson junction $x_\text{qp}^\text{in}$ can be directly inferred from the measured decay rate $\Gamma$ of the transmon's excited state via $\Gamma=Cx_\text{qp}^\text{in}+\Gamma_{0}$. Here $C=\sqrt{2\omega_\text{q}\Delta/\pi^2\hbar}$, where $\omega_\text{q}$ is the frequency of the qubit transition and $\Delta$ is the superconducting energy gap~\cite{catelani2011}. $\Gamma_\text{0}$ is the intrinsic decay rate of the qubit in the absence of laser light, which can be due to a constant background quasiparticle density in the material or other sources of decay such as dielectric loss. In our experiment, we measure the qubit energy relaxation time $T_1 = 1/\Gamma$, from which we extract $x_\text{qp}^\text{in}$ using $C \approx 2\pi\times7.74$~GHz and $\Gamma_0$ measured in the absence of laser light. Measurements using a continuous wave (CW) laser then allow us to observe the effect of absorbed light on the steady-state quasiparticle density, while measurements using a laser pulse allow us to study the recovery dynamics of $T_1$ over the timescales in which quasiparticles are trapped or recombine~\cite{Wang2014}.

\section{Results}

\begin{figure*}[t]
\includegraphics{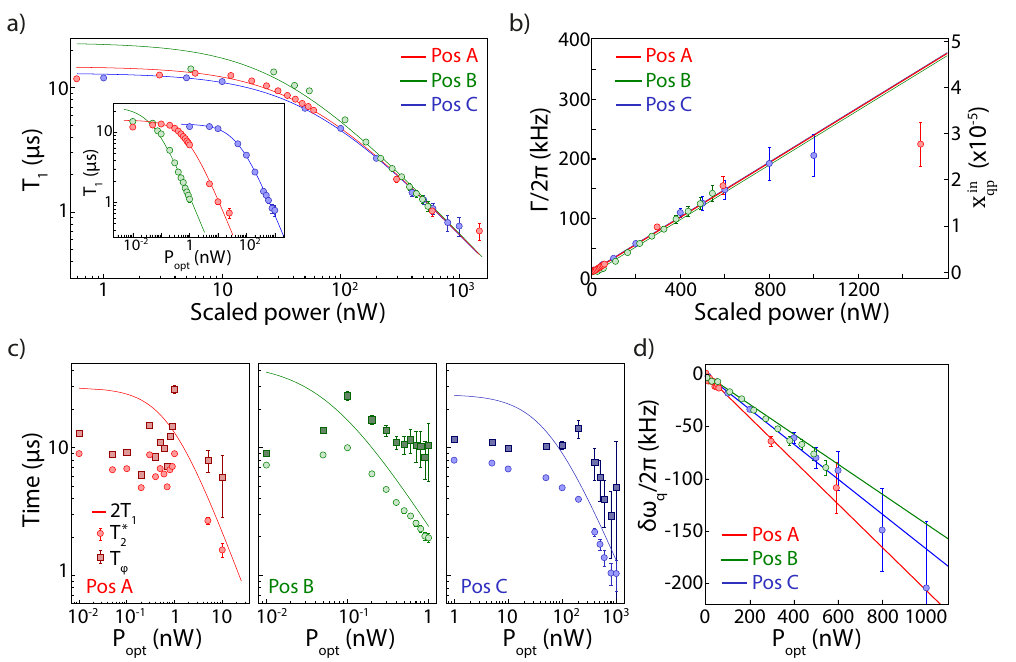}
\caption{\label{fig2} Measurements with CW laser radiation. (a) $T_{1}$ times as a function of laser power, for position A (laser on the qubit), position B (laser at the edge of the qubit pad), and position C (laser far from the qubit). The power axes are scaled for positions A and B by the ratios of the fitted proportionality constants at each position compared to position C (not scaled), i.e., $\mu_\text{A}/\mu_\text{C}\approx 59 $ and $\mu_\text{B}/\mu_\text{C} \approx 544$ respectively, to show their similar behavior. Inset: Same data, without scaling the power axes. (b) Decay rate of the qubit and the corresponding calculated induced quasiparticle density as a function of power for the laser in positions A, B, and C, illustrating the linear dependence of $\Gamma$ with the optical power. Power axes of positions A and B are scaled by the same ratios as in (a). (c) Ramsey measurement of the decoherence time (circles), $2T_{1}$ (solid line), and inferred dephasing time (squares), for the three different positions. We observe that the decoherence times tend to be dominated by decay ($T_2^{*}\sim2T_1)$ for high powers in all three positions. (d) Shift of the qubit frequency from its value with the laser turned off. All three positions show a linear reduction in the qubit frequency, consistent with an increase in $x_\text{qp}^\text{in}$. The power axes are again scaled for positions A and B, using the same ratios as in (a). In (a) and (d), the error bars in each data point represent the standard deviation of ten sequential measurements; in (b) the error bars are propagated from data in (a); in (c), the error bars are calculated using ten sequential measurements and propagation of the error bars in (a).}
\end{figure*}

Measuring the qubit properties in the presence of a CW laser, we find that an increase in the laser power $P_\text{opt}$ incident on the qubit leads to a shorter $T_1$, as shown in Fig.~\ref{fig2}(a). The data is fitted to a model where the light-induced quasiparticle density depends linearly on the optical power, i.e, $x_\text{qp}^\text{in}=\mu P_\text{opt}$, with $\mu$ being a conversion constant between optical power and quasiparticle density. We provide more information about our model in section \ref{sec:SImodel} of the Supplementary Material~\cite{SM}. 

We perform this measurement with the laser located in positions A, B, and C and find a hierarchy of photon-quasiparticle conversion constants given by $\mu_\text{B} = (1.61\pm 0.06)\times 10^{-5}~\text{nW}^{-1}>\mu_\text{A} = (1.75\pm 0.09)\times 10^{-6}~\text{nW}^{-1} >\mu_\text{C} = (2.96 \pm 0.07)\times 10^{-8}~\text{nW}^{-1}$. Position C is the least sensitive to the optical power, which is unsurprising since the laser is focused on the substrate. While most of the light passes through or is specularly reflected at the sapphire-air interfaces and exits through the cavity holes, some diffusely scattered light can remain in the cavity and eventually impinge on the qubit.  We discuss the possible creation of phonons in the substrate in section~\ref{sec:discussion}. Position A shows a higher sensitivity of $T_1$ to the optical power than position C since we are shining the light directly onto the superconducting film. However, since the reflectivity of the Al film at 1550 nm is 97.5$\%$, we believe that most of the light is reflected backwards and leaves the qubit environment through the input hole of the cavity.

Surprisingly, we find that when the laser is focused at the edge of the qubit (position B), $T_1$ is more sensitive to laser power than the other two cases, including position A. We speculate that this is because light incident at the border of the qubit diffracts, leading to more photons scattered toward the qubit or remaining in the 3D cavity instead of exiting through the holes. Therefore, an effectively higher number of photons are absorbed by the superconductor and a more pronounced effect is observed when compared to the other two scenarios. As can be seen in Fig.~\ref{fig2}(b), the qubit decay rate and inferred $x_\text{qp}^\text{in}$ have a linear dependence on the laser power for the three positions, and therefore all three datasets can be collapsed onto a universal curve when the incident power is scaled by the appropriate ratio of photon-quasiparticle conversion constants. 

\begin{figure*}[t]
\includegraphics{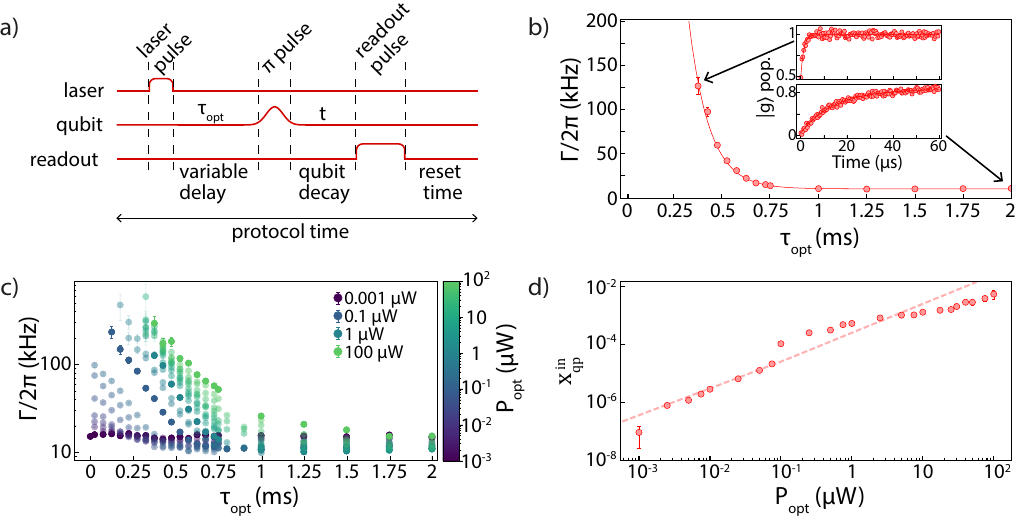}
\caption{\label{fig3} Time-resolved measurements at position A. (a) Protocol for measurement of qubit $T_1$ after a laser pulse. (b) Measurement of the qubit decay rate as a function of variable delay time $\tau_\text{opt}$. Here, we use a $10~\mu$s long laser pulse with an optical power of $1~\mu$W. Insets: Qubit ground state population as a function of time for two different $\tau_\text{opt}$. (c) Qubit decay rate as a function of $\tau_\text{opt}$ for different optical powers. The data for four powers that span five orders of magnitude are highlighted. (d) Initial injected quasiparticle density $x_\text{qp}^\text{in}$ as a function of pulse power, obtained from fitting the data in (c). The dashed line is a linear fit to the data where $P_{opt} \leq$ 0.1 $\mu$W. Error bars in (b) and (c) represent the standard deviation of measurements, while in (d), they represent the propagated error using our model and the estimated trapping and recombination constants.}
\end{figure*}

While quasiparticles have already been demonstrated as an important source of qubit decay, their effects on the decoherence times of the qubits have not been studied as extensively~\cite{lenander2011measurement}. In Fig.~\ref{fig2}(c), we show the effect of light-induced quasiparticles on the $T_\text{2}^*$ decoherence times of our qubit. We can separate the decoherence into two contributions via $1/T_\text{2}^*= 1/ T_{\phi}+1/(2T_\text{1})$,where $T_{\phi}$ is the pure dephasing time. We see that, in the three different positions, $T_\text{2}^*$ decreases with higher optical powers. However, when compared to the $2T_{1}$ curve obtained from the fits in Fig.~\ref{fig2}(a), we see that while the $T_\text{2}^*$ is not $T_1$ limited for smaller powers, it becomes more so for higher powers. We also observe a decrease in the $T_{\phi}$ of the qubit, but the decrease is smaller than that of $T_1$. This indicates that at high powers, the $T_\text{2}^*$ becomes increasingly limited by energy relaxation rather than light-induced quasiparticle dephasing.  

Finally, from the Ramsey measurements, we also extract the quasiparticle-induced qubit frequency shift, $\delta \omega_q$, as a function of the optical power. Fig.~\ref{fig2}(d) shows a reduction in the qubit frequency with higher powers. As described in~\cite{catelani2011}, a reduction in the qubit frequency is another indication of an increase in the quasiparticle density, given the lack of Andreev bound states in a Josephson junction-based transmon. We observe that the frequency shift is negative and linear in the optical power applied, as one would expect from a model of quasiparticle injection (see Equation 73 of Ref.~\cite{catelani2011}). Our linear fits in Fig.~\ref{fig2}(d) yield slopes which are on average 17$\%$ $\pm$ 6$\%$ smaller in absolute value than those expected from theory, when we convert the optical powers to quasiparticle densities. Deviations from the model of this magnitude have been previously observed in Ref.~\cite{Wang2014}. Based on the results of`\cite{Fischer2023}, the smaller slopes could be indicative of quasiparticle heating, but it should be noted that the regime considered there - resonators with a high number of photons below the bandgap - is quite different from the one explored in this work.

After performing experiments with a CW laser, we move to a pulsed regime to study the recovery of qubit lifetime after quasiparticle creation. In Fig.~\ref{fig3}(a), we show the basic protocol we use for this experiment. We use the AOM to create laser pulses with controllable pulse lengths and repetition rates. After the pulse is sent to the qubit, we wait a variable delay time $\tau_\text{opt}$ and perform a $T_1$ measurement. 

A typical measurement result is shown in Fig.~\ref{fig3}(b). Here, we focus on data for the laser located at position A, but similar results are obtained for the laser in the other positions. In the first hundreds of microseconds, we cannot measure the qubit decay rate for higher powers, due to short $T_1$ times~\cite{xu2023lightinduced}. After a certain time that depends on the pulse energy, we are able to measure a $T_\text{1}$ decay curve, as we show in the inset for the first point in Fig.~\ref{fig3}(b). Afterward, we observe a continuous decrease in $\Gamma$ until it reaches an asymptote at $\Gamma_0$. 

We perform this experiment for a large set of optical powers while keeping the optical pulse length fixed at $10~\mu$s. In Fig.~\ref{fig3}(c), we show data for a range of five orders of magnitude in the incident optical power. All plots follow a similar behavior, with an initial higher $\Gamma$, followed by a decline that lasts somewhere between a few hundreds of microseconds to slightly more than one millisecond. Since we use an extensive set of data, taken over several days of measurements, we observe different final asymptotes for the decay rates, which we associate with typical fluctuation observed in the $T_1$ times of the qubit~\cite{Kilmov2018}. To understand the physical origin of the coherence recovery, we fit the data for each power to a model that includes trapping and recombination as possible sources of quasiparticle density decrease. Detailed analysis of the fitting parameters leads us to conclude that we can use a model with a quasiparticle trapping constant of $s=9\pm2$~kHz and no recombination, as discussed in sections \ref{sec:SImodel} and \ref{sec:DataProcessing} of the Supplementary Material~\cite{SM}. This allows us to conclude that we are in a trapping-dominated regime, leading to an exponential recovery of the qubit decay rate

\begin{equation}
    \Gamma(\tau_\text{opt}) = C x_\text{qp}^\text{in}e^{-s\tau_\text{opt}}+\Gamma_0,
\end{equation}

\noindent as we observe in Fig.~\ref{fig3}(b). We use this model to extract the injected quasiparticle density at $\tau_\text{opt}=0$ for each laser power. We show the obtained values in Fig.~\ref{fig3}(d), where we see a clear increase in $x_\text{qp}^\text{in}$ with optical power. At low optical powers, this increase follows a linear trend, while at higher powers the relationship becomes non-linear. Both the optical power and $x_\text{qp}^\text{in}$ span five orders of magnitude, with the latter reaching values as high as $10^{-2}$. 

\begin{figure*}[!ht]
\includegraphics{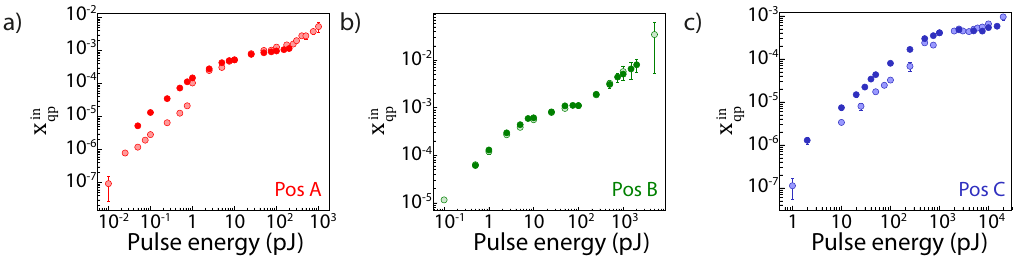}
\caption{\label{fig4} Role of pulse energy. Injected quasiparticle density with laser in (a) position A, (b) position B and (c) position C. In all cases, light points are data for variable optical power, with a fixed pulse length of $10~\mu$s. Darker points are data for variable pulse length, with an optical power of $1~\mu$W (position A), $10~\mu$W (position B), and $100~\mu$W (position C).}
\end{figure*}

Finally, we can use the pulse length as another way to change the total energy injected near the qubit. We expect the effect of changing the optical power and changing the pulse length to be similar, as long as the pulse energy is the same. We compare these effects for the three laser positions, as shown in Fig.~\ref{fig4}. Indeed, we observe qualitatively similar effects for both variations, confirming the role of pulse energy as the determinant control on the injected quasiparticle density. Moreover, we also observe a similar sensitivity hierarchy as the CW experiments, with position C having the lowest photon-quasiparticle conversion constant. However, we observe similar sensitivity values for B and A when taking the error bars into account, which differs from the CW experiment. We note that the powers used in the pulsed experiment are much higher than in the CW case. 

For positions A and C, we also observe that, for some energy ranges, there is a discrepancy between the data where power and pulse length are varied. Furthermore, for all three locations, we find that the dependence of $x_\text{qp}^\text{in}$ on pulse energy is linear at low pulse energies. Slightly below $x_\text{qp}^\text{in}\approx 10^{-3}$, we observe a saturation behavior followed by a further linear increase at the highest energies used. Understanding these effects will require further studies. However, possible explanations could be related to recombination effects at higher quasiparticle densities and diffusion dynamics.   

\section{Discussion and conclusion}
\label{sec:discussion}

In this work, we introduced a technique to create and control quasiparticle dynamics in SC devices. Using an infrared laser, we were able to tune the qubit decay rate by changing the optical energy reaching the device. To rule out a possible temperature increase in the qubit as a source of increased loss, we performed finite element method simulations (see Supplementary Material~\cite{SM} section \ref{sec:SIthermal}), that indicate a negligible increase in the temperature of the device. Furthermore, we find that the excited state population of the qubit is not significantly affected by laser illumination within the range in which it is measurable (see Supplementary Material ~\cite{SM} section \ref{SIepop}). Therefore, we conclude that light-induced quasiparticles are the main source of loss modification in our devices. The increase in losses is directly associated with energy exchange between qubit states and quasiparticle degrees of freedom. 

The strong laser location dependence observed in our experiment indicates that photons directly hitting the SC layer are the most important contribution to quasiparticle creation. This could indicate that processes that lead to quasiparticle generation through the substrate, such as the creation of phonons, do not play a dominant role in sapphire subjected to IR radiation. This observation is particularly important for the field of hybrid systems based on SC qubits, where in many situations one needs to bring light very close to the superconductors~\cite{Mirhosseini2020,Delaney2022}. It is also important to stress that a simple linear model of the absorption of photons, using a photon-quasiparticle conversion constant, allows us to precisely describe the results of our CW experiments. This model could be extended by including higher energy levels of the transmon, which in our model is treated like a two-level system. 

Measurements of $T_\text{2}^*$ also allowed us to observe the behavior of $T_\phi$, which displays a minor reduction in its values for higher powers. The contribution of a thermal distribution of free quasiparticles to $T_\text{2}^*$ through pure dephasing was predicted to be negligible in a single junction transmon when compared to the decay contribution~\cite{CatelaniPRB2012}. We note that certain assumptions of that model, such as a quasi-thermal equilibrium energy distribution with an effective temperature for the quasiparticles, may not hold in our case. Nevertheless, we are able to resolve the effect of quasiparticles on $T_{\phi}$~\cite{lenander2011measurement}, and our results contribute to a better understanding of the quasiparticle-induced decay and dephasing behavior of $T_\text{2}^*$. 

While the CW dynamics can be easily explained by a linear absorption model, the pulsed regime shows more complex dynamics. Nonetheless, using a diffusion-free model with power-independent trapping and recombination rates, we were able to fit all of our results and demonstrate how the injected quasiparticle density depends on the pulse energy. We observe that the qubit coherence recovers with a timescale of  $\tau_{qp} = 0.11\pm0.02$~ms, about one order of magnitude faster than previously reported in an experiment using microwave pulses for quasiparticle injection~\cite{Wang2014}. One explanation for this discrepancy could be that our experiment has a larger stray magnetic field near the qubit, possibly due to the apertures in the microwave cavity. The higher magnetic field would lead to a larger vortex density and result in a larger trapping rate. Alternatively, in our case, highly energetic quasiparticles are created throughout the SC layer rather than at the junction, which could propagate at higher speeds in the material, leading to faster trapping. Finally, we expect to observe a change in the mean free path of the quasiparticles in films with different morphologies, which can result in a variation in trapping rates. However, further studies need to be performed to pinpoint the exact mechanism behind the larger trapping rate. We also note that the quasiparticle lifetime is a material dependent quantity. In this experiment, the relatively long quasiparticle lifetime in aluminum allows us to measure the quasiparticle recovery on the time scale of hundreds of microseconds. For applications where a faster quasiparticle recovery is desired, SC films with much faster quasiparticle recovery times such as niobium and tantalum might be more suitable~\cite{kaplan1976quasiparticle}.

We believe the platform we have developed can be particularly useful to understand \textit{catastrophic} events associated with cosmic ray or environmental radiation absorption in large-scale SC circuits~\cite{McEwen2022}. With the precise control of power, event rate, and location of the events, we can simulate the burst of injected particles in these circuits and evaluate the robustness of a particular circuit to these events. Finally, in future experiments, in-situ control of quasiparticle generation could also be used to study quasiparticle diffusion in SC circuits, contributing to our understanding of how to mitigate their detrimental effects. 

\section*{Acknowledgements}
We thank L. Glazman, A. Grimm, Y. Yang, M. Bild, and T. Schatteburg for valuable discussions. \textbf{Funding:} This project has received funding from the European Research Council (ERC) under the European Union’s Horizon 2020 research and innovation programme (Grant agreement No. 948047). This work was supported by an ETH Zurich Postdoctoral Fellowship. G.C. acknowledges partial support by the U.S. Government under ARO grant W911NF2210257. \textbf{Author contribution:} U.v.L. fabricated the device. R.B, M.D., G.B. and F.A. constructed the setup. R.B. and M.D. performed the measurements and analyzed the data. R.B., M.D., H.M.D., G.C. and Y.C. interpreted the results. R.B., M.D. and Y.C. conceived the experiment. Y.C. supervised the work. R.B., M.D. and Y.C. wrote the manuscript, which was revised by all authors. \textbf{Data and materials availability:} All data and software are deposited on Zenodo~\cite{Zenodo} 

\nocite{*}

\bibliography{QP_paper}

\newpage

\onecolumngrid
\newpage

\begin{center}
    \large{\textbf{Supplementary Material: Quasiparticle dynamics in a superconducting qubit irradiated by a localized infrared source}}
    
\vspace{0.25cm}

   \normalsize R. Benevides$^{1,2}$, M. Drimmer$^{1,2}$, G. Bisson$^{1,2}$, F. Adinolfi$^{1,2}$, U. v. Lüpke$^{1,2}$, H. M. Doeleman$^{1,2}$, G. Catelani$^{3,4}$ and Y. Chu$^{1,2}$

    $^{1}$\textit{Department of Physics, ETH Zürich, 8093 Zürich, Switzerland and}

    $^{2}$\textit{Quantum Center, ETH Zürich, 8093 Zürich, Switzerland}

    $^{3}$\textit{Jülich Aachen Research Alliance (JARA) Institute for Quantum Information (PGI-11), Forschungszentrum Jülich, Jülich 52425, Germany}

    $^{4}$\textit{Quantum Research Center, Technology Innovation Institute, Abu Dhabi 9639, United Arab Emirates}
\end{center}

\appendix

\section{Qubit details}

We use an aluminum 3D transmon with an Al/AlO$_x$/Al junction, fabricated with the use of a single-step, Dolan bridge lithography approach. The typical device properties measured in the absence of laser light in the cavity are shown in table~\ref{tab:tab1}: 

\begin{table}[!h]
    \centering
    \caption{Qubit properties.}
    \begin{tabular}{|c|c|}
    \hline
      Cavity frequency $\text{f}_\text{cav}$  & 9.05 GHz \\ \hline
       Qubit frequency $\text{f}_\text{q}$  & 6.30 GHz \\ \hline
       T$_1$ & 9-11 $\mu$s\\ \hline
       T$_2^*$ & 8-11 $\mu$s \\ \hline
       T$_\phi$ & 14-22 $\mu$s \\ \hline
    \end{tabular}    
    \label{tab:tab1}
\end{table}

\begin{figure*}[!h]
\includegraphics{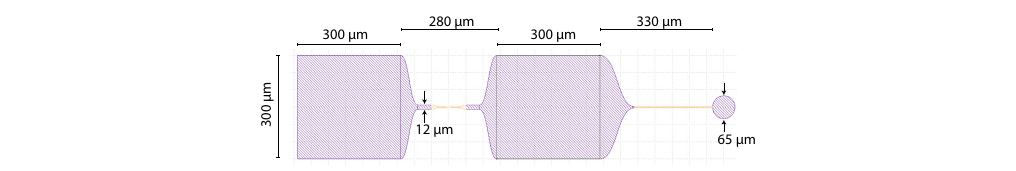}
\caption{\label{fig:fig_si_qubit} Qubit design used in this experiment with the lengths of key features labeled.  }
\end{figure*}

We also present the main qubit design dimensions in Fig.~\ref{fig:fig_si_qubit}. Our qubit is composed of two main pads shunted by a Josephson junction. Additionally, a circular antenna is connected to it and used in other hybrid systems of our group. We shine the laser spot onto the pad at the opposite side of the antenna, to avoid its influence in the results.  

\section{Modelling the experiment}
\label{sec:SImodel}

Previous works have calculated~\cite{catelani2011} and demonstrated~\cite{Wang2014} the relationship between the loss rate of the first excited state of a weakly anharmonic qubit and the normalized quasiparticle density, which is given by

\begin{equation}
\label{eq:basicQPLoss}
    \Gamma = \sqrt{\frac{2\omega_\text{q} \Delta}{\pi^2\hbar}} x_\text{qp}+\Gamma_\text{ext}, 
\end{equation}

\noindent where $\Gamma_\text{ext}$ is qubit relaxation rate due to non-quasiparticle related sources. Here, we expand this analysis to include possible modifications in this relationship when shining a CW laser on the sample. 

\subsection{Qubit frequency modification}

We observe a qubit frequency shift with power in Fig.~\ref{fig2}(d) of the main text, which would modify the observed loss rate as

\begin{equation}
    \omega_\text{q}=\omega_\text{q}^{0}-\lambda P_\text{opt},
\end{equation}

\noindent where $\lambda$ is a proportionality constant. However, we note that the observed shift is much smaller than the absolute frequency ($\lambda P_\text{opt}/\omega_\text{q}^{0}<< 1)$, so we can neglect this contribution in our analysis.

\subsection{Energy gap modification}

For temperatures $T<<T_\text{c}$, where $T_\text{c}$ is the critical temperature of the material, the energy gap would be modified as

\begin{equation}
    \Delta_\text{qp}\approx \Delta(1-x_\text{qp}).
\end{equation}

Considering that we only observe values of quasiparticle density lower than $x_\text{qp}\approx 10^{-2}$, we consider in our model that the energy gap of the superconductor will remain constant during the experiment, neglecting any modification in the energy gap with laser power.

\subsection{Photon-quasiparticle conversion}

We assume that the number of created quasiparticles is proportional to the number of incident photons, i.e., 

\begin{equation}
\begin{split}
    x_\text{qp}&=x_{0}+x_\text{qp}^\text{in}\\
    &=x_{0}+\mu P_\text{opt},
\end{split}
\end{equation}

\noindent where $x_\text{0}$ is the steady-state quasiparticle density in the absence of quasiparticles injected by laser light and $\mu$ the photon-quasiparticle conversion constant, which is position dependent. We disregard any non-linear behavior and two-photon processes in the dynamics of conversion. 

\vspace{0.5cm}

Therefore, the full model that relates CW optical power to qubit decay rate is given by 

\begin{equation}
    \Gamma=\sqrt{\frac{2\omega_\text{q} \Delta}{\pi^2\hbar}}\mu P_\text{opt}+\Gamma_{0},
\end{equation}

\noindent where $\Gamma_0=\Gamma_\text{ext}+\sqrt{2\omega_\text{q} \Delta/\left(\pi^2\hbar\right)}x_{0}$ is the qubit relaxation rate in the absence of quasiparticle injection which includes the contribution of the background quasiparticle density and other sources of loss such as dielectric loss. 

\subsection{Quasiparticle recovery model}\label{sec:QPRecoveryModel}

Next, we proceed to understand the temporal behavior of injected quasiparticles. If we assume that the quasiparticle diffusion has much faster timescales than recombination and trapping~\cite{Wang2014}, diffusion can be neglected (see section~\ref{sec:DataProcessing}) and the following equation models the quasiparticle density~\cite{rothwarf1967measurement}

\begin{equation}
\label{eq:QP_EOM}
    \frac{d x_\text{qp}}{d t} = - r x^2_\text{qp} - s x_\text{qp} + g
\end{equation}

\noindent where $r$ and $s$, are the quasiparticle recombination and trapping, respectively, and $g$ is the quasiparticle generation rate in the absence of laser light. The solution of this equation is

\begin{equation}
\label{eq:QPdensitySolution}
    x_\text{qp}(t) = \frac{ x_\text{qp}^\text{in} ( s + 2 r x_0) }{ ( s + r x_\text{qp}^\text{in} + 2 r x_0) e^{ ( s + 2 r x_0 ) t } - r x_\text{qp}^\text{in} } + x_0
\end{equation}

\noindent where $x_\text{qp}^\text{in}$ is the initial injected quasiparticle density and $x_0$ is the steady-state quasiparticle density (specifically, the steady-state solution of Eq~\ref{eq:QP_EOM}). The quasiparticle density influences the decay rate of the qubit according to the equation~\ref{eq:basicQPLoss}, so we can now write down the expression for the qubit relaxation rate in terms of $r$, $s$, $x_\text{qp}^\text{in}$, and $\Gamma_0$. To simplify the expression, we neglect qubit losses from other sources and let $\Gamma_0 = \sqrt{2\omega_\text{q}\Delta/\pi^2\hbar} x_0$, meaning that the solution only puts an upper bound on the background quasiparticle density. With this assumption, the qubit decay rate  can be written as

\begin{equation}
\label{eq:QPrecoveryModel}
    \Gamma(t) = \frac{ C x_\text{qp}^\text{in} \left( C s + 2 r \Gamma_\text{0}\right) }{ \left( C s + C r x_\text{qp}^\text{in} + 2 r \Gamma_\text{0}\right) e^{ \left( s + \frac{2 r \Gamma_\text{0} }{C} \right) t } - C r x_\text{qp}^\text{in}  } +\Gamma_\text{0}
\end{equation}

\noindent where we define, as in the main text, $C = \sqrt{2\omega_\text{q}\Delta/\pi^2\hbar}$ to simplify notation.

\section{Details of the experimental setup}
\label{sec:SI-details}

\begin{figure*}[!b]
\includegraphics{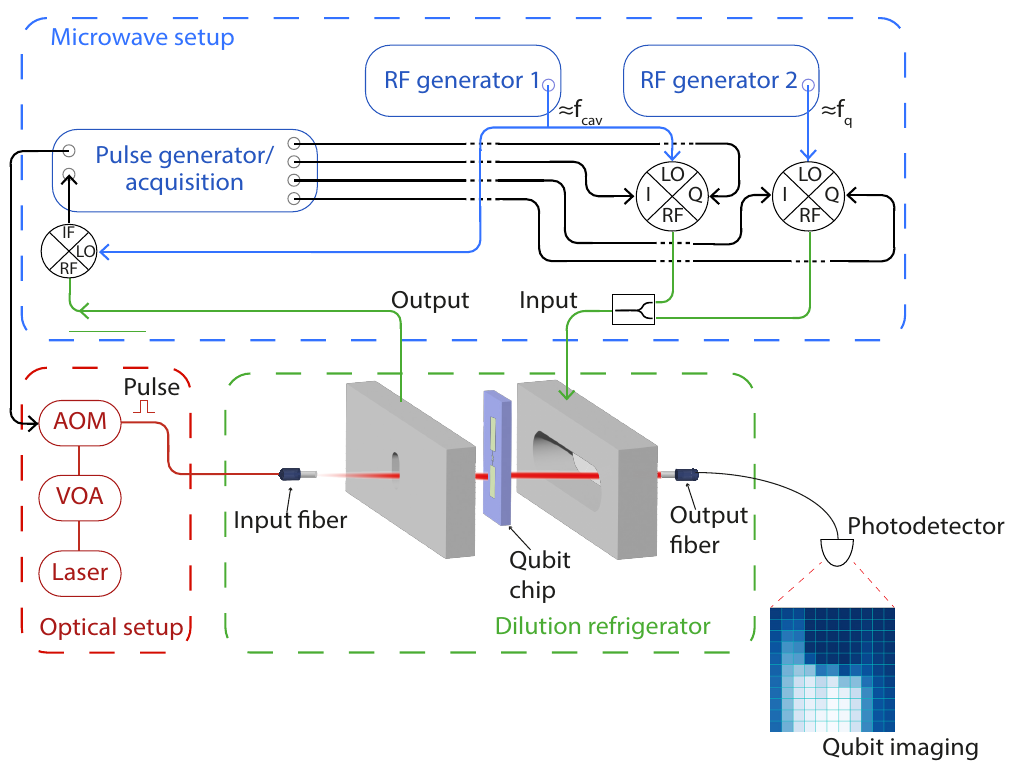}
\caption{\label{fig:SI2}Experimental setup. A standard circuit QED setup is used to operate and read our qubit. Scanning the position of the laser with the help of a cryo-actuator and collecting the transmitted power, we can image the qubit and locate the laser spot. VOA: variable optical attenuator, LO: local oscillator, RF: radiofrequency, IF: intermediate frequency.}
\end{figure*}

\begin{figure*}
\includegraphics{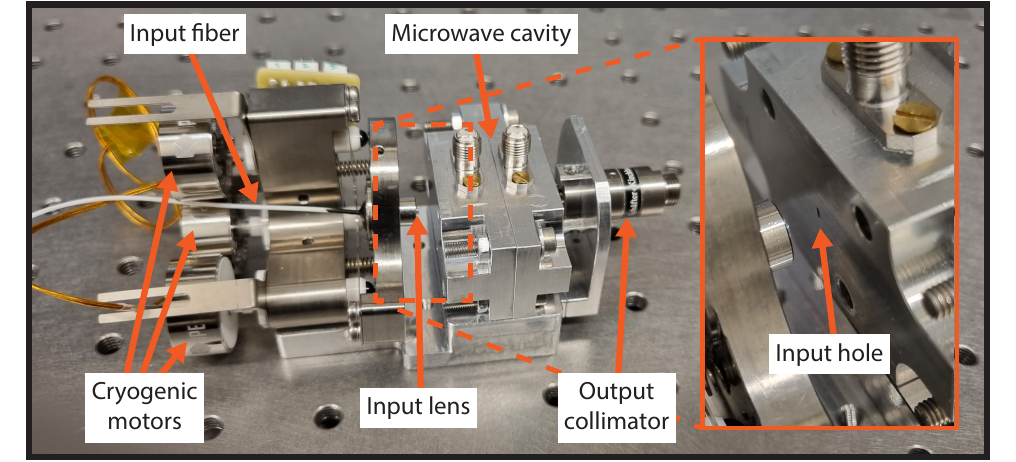}
\caption{\label{fig:fig_si3} Photo of the experimental device. Light enters from the optical input fiber on the left and is focused by the input lens through the input hole onto an SC qubit inside the microwave cavity. The inset shows a zoomed-in alternate angle of the input hole to the microwave cavity. If the light is transmitted through the cavity, it can exit through an output hole and be collected by the output collimator on the right side. The cryogenic motors can move the tip-tilt stage and control the position of the optical beam at the SC qubit.}
\end{figure*}

The qubit measurements are performed using a standard circuit QED setup, as can be seen in Fig.~\ref{fig:SI2}. The qubit is dispersively read-out using the 3D microwave cavity.  An FPGA-based quantum control unit (Quantum Machines OPX) and IQ-mixers are used to create microwave pulses for addressing the qubit and readout cavity. The optical pulse is controlled via a tunable laser kept at $\lambda=1550$~nm, a variable optical attenuator (VOA),  and an acousto-optical modulator (AOM). The AOM is also driven using RF pulses created by the control unit, allowing the synchronization of the full protocol. The laser light is sent via an optical fiber into the fridge and outcoupled using a graded-index (GRIN) lens attached to a motorized actuator (Janssen Precision Engineering CTTPS1/2) to control the beam position (see Fig.~\ref{fig:fig_si3}). An output collimator collects the transmitted light and allows for in-situ imaging of the sample. 

\section{Imaging inside the dilution refrigerator}
\label{sec:SI-imaging}

\begin{figure*}
\includegraphics{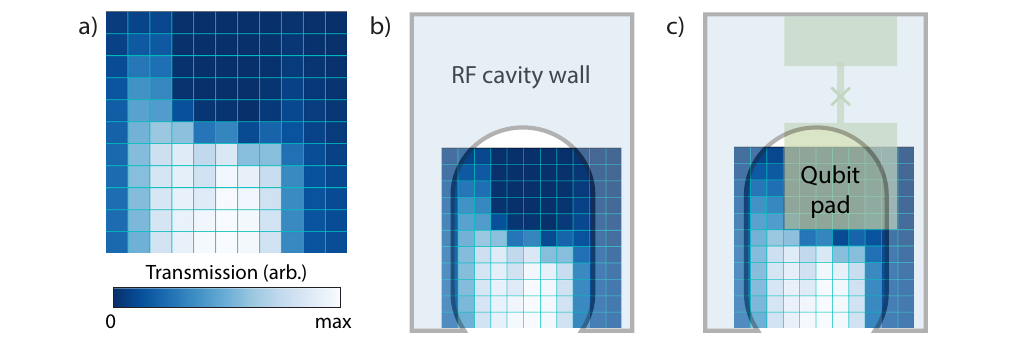}
\caption{\label{fig:fig_si4} Imaging protocol and interpretation. (a) A crude picture of the qubit is taken inside the fridge with the help of the laser and cryogenic motors. (b) We associate the continuous straight lines at the sides of the image with microwave cavity walls and (c) the square region with lower optical transmission with the qubit lower pad. }
\end{figure*}

One experimental challenge we faced in performing the experiments described in this work was positioning the laser spot with respect to the superconducting circuit in a dilution refrigerator. We developed a crude imaging technique in order to get information about the laser beam's location. By moving the tip-tilt stage of the input GRIN lens in fixed steps and monitoring the transmitted laser light collected by the output collimator, we can create an image of the qubit in the cavity. One such image is shown in Fig.~\ref{fig:fig_si4}(a).

There are two relevant features in these images. Firstly, when the laser beam is blocked by the edge of the holes in the microwave cavity, no light is transmitted. Therefore, we interpret vertical features that span the entire image as the walls of the microwave cavity (see Fig.~\ref{fig:fig_si4}(b)). Secondly, when the laser spot is incident on the rectangular piece of aluminum that makes up the bottom qubit pad, the light is reflected and we measure no transmission (see Fig.~\ref{fig:fig_si4}(c)). Once we take and interpret an image, we use the motors to choose a position for the laser beam spot with respect to the qubit.

\section{Thermal analysis with finite element methods}

\label{sec:SIthermal}

\begin{figure*}[!b]
\includegraphics{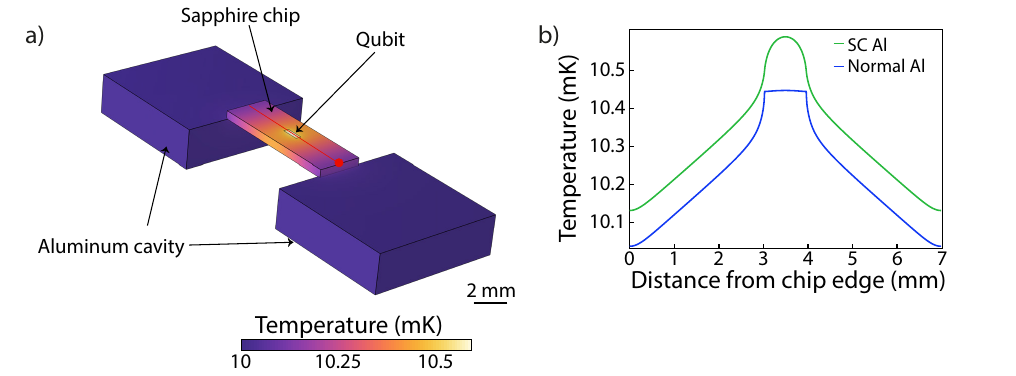}
\caption{\label{fig:fig_si5} Simulation of the sample temperature. (a) Temperature distribution in the sample and aluminum blocks representing the cavity, with the aluminum assumed to be in its superconducting phase. We use a rectangular aluminum layer to represent the qubit. The red line represents the line-cut of data shown in (b), with the origin of the distance axis at the red circle.} 
\end{figure*}

One potential complicating factor in this experiment is the effect of the laser on the temperature of the qubit sample. Even short low-power pulses could potentially increase the temperature of the fridge and modify the thermal quasiparticle background. To check this effect, we performed finite element methods simulation to compute the expected increase in temperature in our device. We modeled our setup as a sapphire chip supported by blocks of aluminum (representing the cavity) and with a rectangular layer of aluminum in its surface with the approximate dimensions of the qubit, as shown in Fig.~\ref{fig:fig_si5}(a). We assumed that the precise qubit geometry was irrelevant to the analysis, so we ignored it to simplify the meshing. We also assumed the cavity is in thermal contact with a bath at the fridge temperature of 10 mK. During the experiment, we observed that the temperature of the fridge reaches a maximum of $\sim35$~mK for the highest powers in the CW experiment and is not significantly affected for small powers or the pulsed experiment.

We modeled the effect of our laser considering the situation where the highest CW power was sent to the experiment, i.e., in position C with $P_\text{opt}=1~\mu$W. We consider a typical absorption of infrared light by aluminum layers of $2.5\%$ and that in position C we have a reduction of $\mu_\text{A}/\mu_\text{C}\approx 59 $ of the power reaching the qubit. We used thermal conductivities of $0.002~$W/(m$\cdot$K) for the sapphire and the aluminum in the superconducting phase at $100$~mK~\cite{pobell2013matter}. Since the absorption of laser light could lead to a local transition in the superconductor material to the normal phase, we considered also the situation where the metal would fully transition out of the superconducting phase, using the value of $200~$W/(m$\cdot$K).

We computed the stationary temperature of the whole system under the illumination of the laser, as shown in Fig. \ref{fig:fig_si5}(a). In Fig. \ref{fig:fig_si5}(b), we show a cut-line of the temperature across the qubit chip surface. We observed an increase no bigger than $0.6$~mK above the fridge base temperature for this configuration. We have also analyzed the effect of a higher temperature bath (fridge temperature) and of different positions of the laser (with the respective used optical powers). In all cases, we didn't see changes in the temperature that could explain the measured quasiparticle densities. We note that assumptions made for this simulation, such as the perfect thermal contact between materials, the lack of absorption in the sapphire and the uncertainty about the thermal conductivities, limit the precision of the obtained results. Still, considering the linearity of the heating with respect to absorbed optical power, these results provide one piece of evidence that heating of the qubit does not play an important role in our experiments, consistent with the measurements observed in the next section.  

\section{Measurements of the excited state population of the qubit}\label{SIepop}

To further investigate the possible heating of the qubit chip by the laser, we performed qubit thermometry measurements. We compared the amplitude in the power Rabi oscillation of the $e \leftrightarrow f$ transition when preceded by a $\pi$-pulse between the $g \leftrightarrow e$ states with the same power Rabi without the initial $\pi$-pulse. Assuming negligible population in the $f$ state, this method provides the excited state population of the qubit~\cite{Geerlings2013}, which translates to an effective temperature assuming a Boltzmann distribution. 

\begin{figure}[!h]
\includegraphics{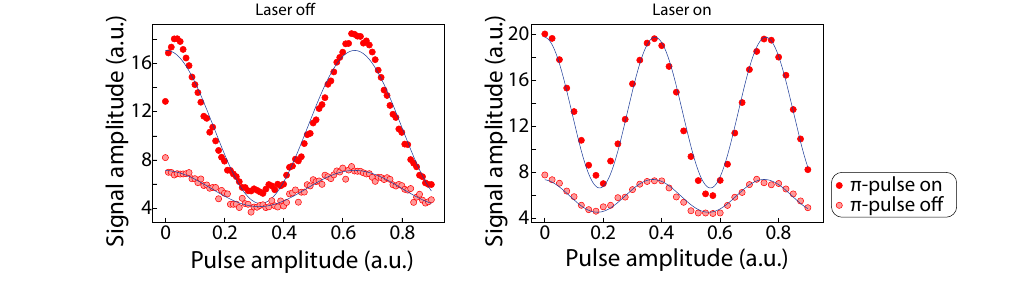}
\caption{\label{fig:fig_si6} Qubit temperature measurement. Rabi oscillations between states $e \leftrightarrow f$, with laser off and with CW laser on in position A and $P_\text{opt} = 1$~nW, both with and without an initial $\pi$-pulse between states $g \leftrightarrow e$.}
\end{figure}

In Fig.~\ref{fig:fig_si6}, we show such a measurement, performed both with the laser off and with the laser on at position A with a CW power of $P_\text{opt} = 1$~nW. We observe in both situations similar temperatures of $T\approx 160-165$~mK, considerably higher than our fridge base temperature ($\sim 10$~mK) or our estimations based on finite element methods (section~\ref{sec:SIthermal}), but still much lower than what is necessary to explain quasiparticle densities observed in the experiment, which reach $x_\text{qp}\sim 10^{-2}$. We associate these higher temperatures with imperfect thermalization to the base plate and with black-body radiation entering the microwave cavity through its hole, which is known to contribute to the transition rates of the qubit via photon-assisted tunneling and breaking of Cooper pairs~\cite{Houzet2019}. Therefore, from the measurements performed, we believe the optical pulses are not significantly heating up the qubit. We summarize the measured values in the table~\ref{tab:tab2}, with the excited state population, the qubit temperature, and the corresponding thermal quasiparticle density $x_\text{qp}^\text{th}$~\cite{Glazman2021} for each of the situations presented.

\begin{table}[!h]
    \centering
    \caption{Thermometry of the qubit with and without laser.}
    \begin{tabular}{|c|c|c|c|}
    \hline
         & $|e\rangle$ population & Qubit temperature (mK) & $x_\text{qp}^\text{th}$ \\ \hline
       Laser off & $0.191\pm 0.006$ & $\sim 165$ & $\sim 7.8\times10^{-7}$\\ \hline
       Laser on & $0.178\pm 0.007$ & $\sim 160$ & $\sim 5\times10^{-7}$\\ \hline
    \end{tabular}
    \label{tab:tab2}
\end{table}

\section{Quasiparticle relaxation as a function of pulse length}

\begin{figure}
\includegraphics{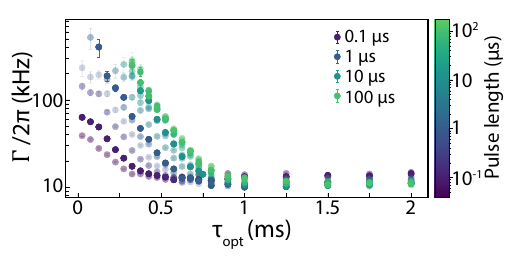}
\caption{\label{fig:fig_si7} Pulsed laser at position A, pulse length sweep. Qubit decay rate as a function of $\tau_\text{opt}$ for different optical pulse lengths. The data for four pulse lengths that span four orders of magnitude are highlighted. Error bars represent the standard deviation of measurements.}
\end{figure}

In Fig.~\ref{fig3} of the main text, we investigated quasiparticle relaxation after a pulsed laser experiment while changing the optical power and keeping the pulse length fixed. Here, we show in Fig.~\ref{fig:fig_si7} a qualitatively similar behavior when changing the optical pulse length and keeping the optical power fixed at $10~\mu$W. We observe quasiparticle relaxation on similar timescales. Nevertheless, both cases allow us to control the injected quasiparticle density across 4 to 5 orders of magnitude. The data in Fig.~\ref{fig:fig_si7} were taken with the laser at position A. These data (as well as similar data taken at positions B and C) are analyzed and shown in Fig.~\ref{fig3}.

\section{Details on data processing}
\label{sec:DataProcessing}

In this section, we will explain the procedure for data acquisition, fitting, and analysis from the quasiparticle recovery measurements shown in Fig.~\ref{fig3}(b)-(d). Our data starts as a series of $T_1$ measurements corresponding to a particular laser pulse power ($P_\text{opt}$), pulse length ($\tau_\text{pulse}$), and the time delay between the laser pulse and the measurement ($\tau_\text{opt}$). For each series of such measurements, we fit the $T_1$ measurement to an exponential model and obtain a qubit energy relaxation rate, $\Gamma(P_\text{opt},\tau_\text{pulse},\tau_\text{opt})$, and a standard error of the fit $\sigma(P_\text{opt},\tau_\text{pulse},\tau_\text{opt})$, which correspond to the data point and error bar of each point in Fig.~\ref{fig3}(b). 

When the pulse energies are very high and the time delays between the pulse and the measurement are short, the $T_1$ measurements can become too noisy to fit. When the signal to noise becomes too low, the fitted $T_1$ becomes larger when $\tau_\text{opt}$ is decreased and the fit error increases. If we observe this unphysical effect, we identify the pulse delay $\tau_\text{opt}$ after which the qubit loss rates decrease monotonically and only consider data with a delay greater than or equal to that $\tau_\text{opt}$. 

For a given laser pulse power and pulse length, we then fit the qubit loss rates to our model of quasiparticle recovery, Eq.~\ref{eq:QPrecoveryModel}. In this model, there are two free parameters, the initial injected quasiparticle density $x_\text{qp}^\text{in}$ and the qubit loss rate in the absence of injected quasiparticles $\Gamma_0$. There are also two unknown constants, the recombination and trapping rates $r$ and $s$. If we leave $x_\text{qp}^\text{in}$, $\Gamma_0$, $r$, and $s$ as free parameters, the magnitude of the cross-correlations between variables in the fit approach unity and the fit becomes unreliable, so we need to fix $r$ and $s$. We know that for low enough quasiparticle densities (when $x_\text{qp} \ll s / r $), trapping effects will dominate over the recombination effects. Using literature values~\cite{Wang2014}, we know that recombination and trapping rates should be on the order of MHz and kHz, respectively, so we first fit the data for powers less than 100 nW (where $x_\text{qp}^\text{in} \leq 10^{-4}$), where we enforce $r=0$. These fits yield a more reliable estimate of $s$. From the average and standard deviation of the fitted trapping rates, we find $s = 9 \pm 2$~kHz. Having estimated $s$, we can compare $1/s\approx 0.11$~ms to the diffusion time. Using the diffusion constant in~\cite{Wang2014} and the qubit geometry, we can estimate that the diffusion time across the pad is about $0.05$~ms, less than half of the trapping time. So, neglecting diffusion is a reasonable approximation.

We then fix this value for $s$ and attempt to find a value for the recombination rate $r$ by simultaneously fitting the qubit recovery data for all of the powers we measured. We use the average $\chi^2$ goodness of fit parameter to assess how varying $r$ between 0 and 20 MHz affects the model. We found that  $\chi^2$ was minimized for $r=0$, and the fit results have a very weak dependence on $r$, with a change of $\leq 5\%$ in the values of $\chi^2$ for $r\leq 4.86$~MHz. Therefore we set the recombination rate to zero in our quasiparticle recovery model. With the trapping and recombination rates fixed, we finally fit our $T_1$ recovery data (i.e. Fig.~\ref{fig3}(c)), weighted by the measured variance of each qubit relaxation rate, to ascertain the quasiparticle relaxation behavior (i.e. Fig.~\ref{fig3}(d)).

\end{document}